\documentclass[
reprint,
amsmath,
prl,
superscriptaddress,
noeprint,
]{revtex4-1}
\usepackage{siunitx}
\usepackage{graphicx}
\usepackage{dcolumn}
\usepackage{braket}
\usepackage{xfrac}
\usepackage{hyperref}
\usepackage[usenames,dvipsnames]{color}
\usepackage{tikz-cd}
\usepackage{float}
\usepackage[nameinlink,capitalise]{cleveref}
\usepackage{mathtools}
\usepackage{amsmath}
\usepackage{ulem}
\DeclareSIUnit\gauss{G}
\DeclareSIUnit{\au}{{a.u.}}
\hypersetup{
colorlinks=true,
linkcolor=blue,
citecolor=blue,
filecolor=green,
urlcolor=blue,
}

\begin{document}
\title{Universal probability distributions of scattering observables in ultracold molecular collisions}

\author{Masato Morita}
\affiliation{Department of Physics, University of Nevada, Reno, NV, 89557, USA}
\author{Roman V. Krems}
\affiliation{Department of Chemistry, University of British Columbia, Vancouver, British Columbia V6T 1Z1, Canada}
\author{Timur V. Tscherbul}
\affiliation{Department of Physics, University of Nevada, Reno, NV, 89557, USA}

\date{\today}
\begin{abstract}
Currently, quantum scattering calculations cannot be used for quantitative predictions of molecular scattering observables at ultralow temperatures. This is a result of two problems: the extreme sensitivity of the scattering observables to details of potential energy surfaces (PES) for interactions between the collision partners, and the exceedingly large size of molecular basis sets required for the numerically exact integration of the Schr\"{o}dinger equation in the presence of external fields.  Here, we suggest a new statistical approach to address both of the above problems. We show that ensembles of scattering calculations with different PESs are characterized by cumulative probability distributions, which are insensitive to the size of the molecular basis sets and can, therefore, be obtained from calculations with restricted basis sets. 
This opens the possibility of making predictions  of experimentally relevant observables for a wide variety of molecular systems, currently considered out of reach of quantum dynamics theory. 
We demonstrate the method by computing the success probability of sympathetic cooling of CaH and SrOH molecules by Li atoms and SrF molecules by Rb atoms. 
\end{abstract}

\maketitle

Understanding and controlling the quantum  dynamics of ultracold molecular collisions in the presence of external electromagnetic fields is a central goal of ultracold molecular physics \cite{Rev_Carr08,Rev_Bohn17,Moses:17, Krems:18}, a rapidly evolving research field with a wide range of applications in physical chemistry \cite{Rev_Krems08},  quantum information processing \cite{DeMille:02}, quantum simulation \cite{Rev_Bohn17,Moses:17}, and  precision tests of   fundamental physics beyond the Standard Model~\cite{Rev_Carr08,Rev_Krems08,Rev_Bohn17,Rev_Safronova18}. Recent highlights include the production of high phase-space-density ensembles
of polar radicals SrF($^2\Sigma$) and CaF($^2\Sigma$) via laser cooling and  magnetic trapping \cite{DeMIlle_PRL18,Anderegg_Nature18},  fundamental studies of ultracold molecular collisions and chemical reactions in the single partial-wave regime \cite{Klein:16,Perreault:17}, external field control of reaction dynamics \cite{Ni_NATURE10,Rev_Bala16} and stereodynamics  \cite{Miranda:11}, the observation of dipolar exchange interactions between ultracold KRb molecules trapped in an optical lattice \cite{Yan:13}, and order-of-magnitude better limits on the electric dipole moment of the electron \cite{Andreev:18}.

\begin{figure}[t]
\includegraphics[width=1.0\linewidth]{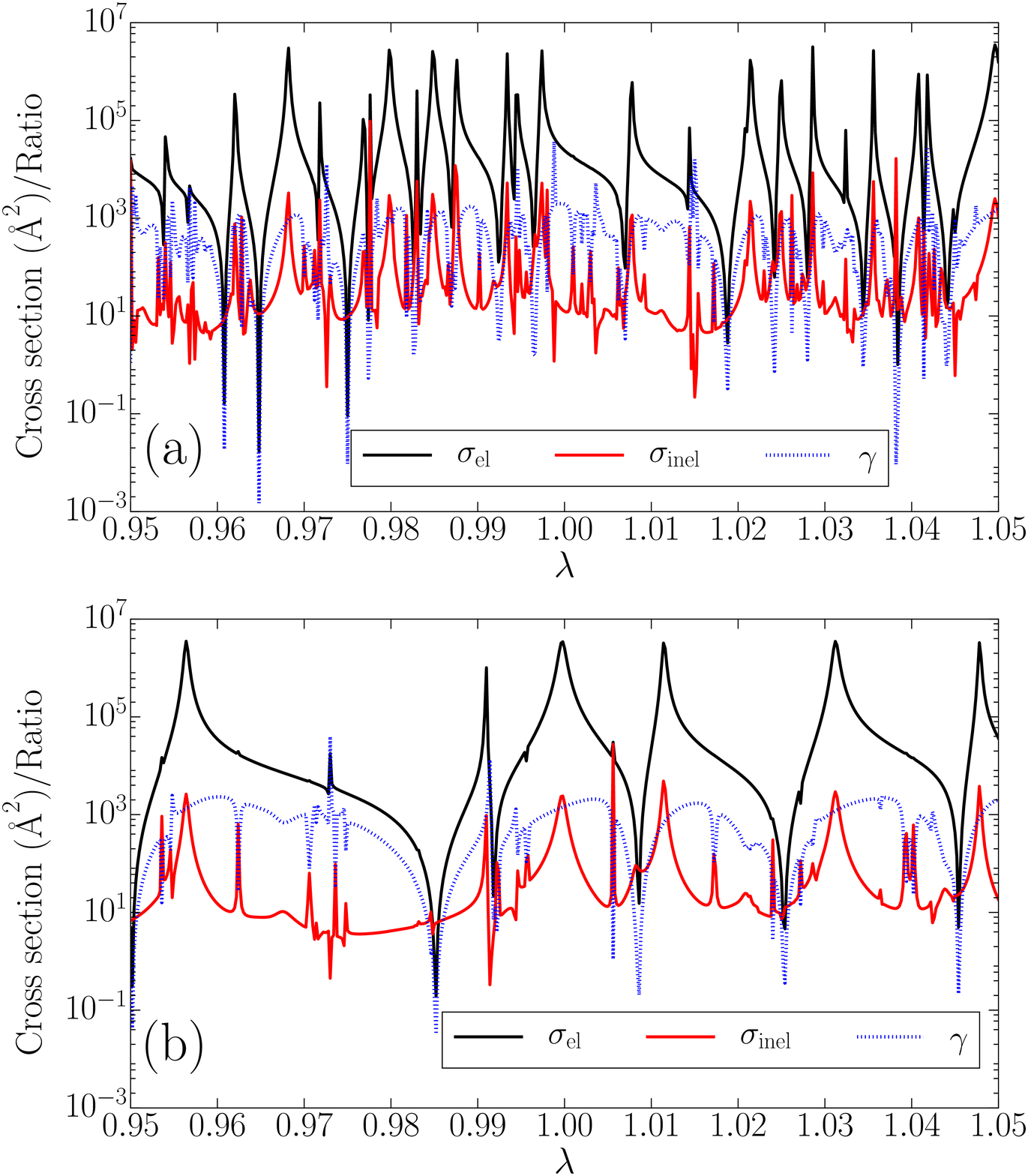}
\caption{Elastic cross section $\sigma_\text{el}$ (black solid), inelastic cross section $\sigma_\text{inel}$ (red solid) and the ratio of elastic-to-inelastic cross sections $\gamma$ (blue dashed) as a function of the potential scaling parameter $\lambda$ for Li-CaH at the collision energy of $E_C=10^{-5}$ cm$^{-1}$ with the external magnetic field of 100 G for $N_\text{max}=55$ (a) and $N_\text{max}=5$ (b). }
\label{figure1}
\end{figure}


As molecular phase-space densities continue to increase toward quantum degeneracy, molecular collisions will play an increasingly important role in ultracold molecular physics. First and foremost, understanding ultracold collisional mechanisms is essential for controlling molecular interactions with external electromagnetic fields \cite{Lemeshko:13}, which is a necessary ingredient in virtually any application of ultracold molecular gases \cite{Rev_Carr08}.
Using external fields to control chemical  reactivity at ultralow temperatures is a major thrust of ultracold controlled chemistry \cite{Rev_Krems08}.
Finally, collisions form the basis of a powerful technique of sympathetic cooling, whereby a trapped molecular ensemble is allowed to thermalize with an ultracold atomic gas. While elastic atom-molecule collisions contribute to cooling, inelastic collisions lead to undesirable trap loss and heating. 
A large ratio of elastic to inelastic collision rates ($\gamma\ge 100$) is a requirement for efficient cooling \cite{Rev_Carr08}. 
 Quantum scattering calculations of molecular collisions are required to guide experiments in the choice of molecular systems amenable to sympathetic cooling. However, at present, quantum scattering calculations cannot be used for quantitative predictions of molecular collision properties at ultracold temperatures, for two reasons.

First, low-temperature scattering observables are extremely sensitive to small uncertainties in the potential energy surface (PES) underlying quantum scattering calculations. 
This well-known phenomenon is illustrated in Figure 1, showing the variation of scattering cross sections with a scaling factor modifying the PES for a relatively simple Li - CaH collision system. 
As can be seen, a small variation ($< 1 \%$) of the PES may change the scattering cross sections by as much as ten orders of magnitude. This characteristic phenomenon is a result of the presence of a large number of scattering resonances which emerge at zero energy as the potential is scaled. The uncretainty of the PES produced by current quantum chemistry methods, especially for heavy open-shell systems of relevance to the research field of ultracold molecules, is much larger than 1 \%.

The second problem making quantum predictions of ultracold scattering observables unfeasible is the Hilbert space dimensionality problem. 
In order to obtain numerically exact quantum results for molecular collisions at ultracold temperature, it is necessary to account for the enormous number of rotational, fine, and hyperfine molecular states coupled by strongly anisotropic intermolecular interactions. This is exacerbated by the need to account for the couplings induced by external fields, which are necessarily larger than the kinetic energy of the colliding molecules, and which therefore spoil the conservation of the total angular momentum. 
 The number of quantum states participating in ultracold molecular dynamics at ultracold temperatures is thus prohibitively large. If the Hilbert space of molecular states (basis set for molecular scattering calculations) is reduced to a feasible size, the scattering observables exhibit the same sensitivity to the number of molecular states as to the interaction potential.

In this Letter, we suggest a new statistical approach to address both of the above problems. We argue that the results displayed in Figure 1 contain meaningful statistical information, which can be used to predict the probability of scattering events. In particular, we show that an ensemble of scattering results with different PES could be characterized by the cumulative probability distributions (CPDs). Moreover, we show that such CPDs appear to be universal to different basis sets and can thus be obtained with extremely high precision based on scattering calculations with small basis sets.  
This is critically important as this opens the possibility of making predictions  of experimentally relevant observables for a wide variety of molecular systems, currently considered out of reach of quantum dynamics theory. 


As a first application, we calculate the CPDs for  cold Li-CaH, Li-SrOH and Rb-SrF collisions in a magnetic field to estimate  the success probability of sympathetic cooling of $^2\Sigma$ molecules with the alkali-metal atoms in a magnetic trap. We show that the CPDs of the elastic-to-inelastic ratio $\gamma$ provide a natural way to estimate the success probability, leading an efficient  methodology for screening  molecular candidates with favorable collisional properties for atom-molecule sympathetic cooling.

We begin by defining the CPD of a scattering observable $\gamma$, which we assume to be a random variable drawn from an ensemble $\left\{{\gamma_{i}} \right\}$  (in the following $\gamma$ will be identified with the elastic-to-inelastic  ratio) 
\begin{equation}\label{eq:CP}
F(\Gamma) = P(\gamma\le\Gamma),
\end{equation}
where $P(\gamma\le\Gamma)$ is the probability that $\gamma$ does not exceed  $\Gamma$. In the limit of infinitely large ensemble size, the CPD may be expressed as $F(\Gamma) = \int_{0}^{\Gamma} p(\gamma)d\gamma$, where $p(\gamma)$ is the probability density function of the observable $\gamma$.
Assuming that an atom-molecule collision pair is suitable for sympathetic cooling  if $\gamma>100$ \cite{Rev_Carr08}, we can define the \textit{success probability} of sympathetic cooling as $P_s=1-F(\Gamma=100)$, which represents the fraction of the elements in the ensemble for which $\gamma>100$.

To obtain the statistics of scattering observables necessary to evaluate the CDF,
we  sample the interaction PES $V$ using the potential scaling method, whereby the atom-molecule PES is multiplied by a dimensionless scaling factor $\lambda$, which is varied to introduce the PES uncertainty
\cite{Hutson_PRL09, Hutson_EPJD11, Hutson_PCCP11, Krems_PRA13, Suleimanov_JPB16}. 
We scan the value of $\lambda$ from 0.95 to 1.05 with the grid step $\Delta \lambda=2\times 10^{-4}$, leading to
a sample with the size of 501, \{$V_i$\} ($ i=1,2,...,501$), 
drawn from the ensemble with $\pm$ 5\% uncertainty, which corresponds to a typical accuracy of modern \textit{ab initio} PESs \cite{Hutson_PRL09}. For $V$, we use the accurate {\it ab initio} interaction PESs  for Li-CaH, Li-SrOH, and Rb-SrF  from our previous work \cite{Tscherbul_PRA11,Morita_RbSrF,Morita_PRA17}. We note that the precise estimate of the uncertainty is not important for our statistical approach as shown in the Supplemental Material \cite{SM}.   
By carrying out CC calculations for every $V_i$ in the sample \{$V_i$\}, we obtain samples of scattering observables such as the elastic $\left\{  {\sigma_{\mathrm{el},i}} \right\}$ and inelastic $\left\{  {\sigma_{\mathrm{inel},i}} \right\}$ cross sections and their ratio $\left\{\gamma_{i}  \right\}$.
We note that more elaborate random sampling methods produce essentially the same CPDs as the standard $\lambda$-scaling method used here \cite{SM}. 

Having specified the potential ensembles, we proceed to carry out numerically exact quantum scattering calculations on ultracold collisions of  
alkali-metal atoms (A) with $^2\Sigma^+$ radicals (B)  in the presence of an external magnetic field. 
In atomic units, the Hamiltonian of the collision complex is \cite{Morita_RbSrF,Morita_PRA17}
\begin{equation}
\fontsize{8.5pt}{0pt}\selectfont \hat{\mathcal{H}} = - \frac{1}{2\mu R}  \frac{d^2}{dR^2} R
                     + \frac{(\hat{J}-\hat{N}-\hat{S}_{\mathrm{A}}-\hat{S}_{\mathrm{B}})^2}{2\mu R^2} \allowbreak
                     + \hat{\mathcal{H}}_{\mathrm{A}}
                     + \hat{\mathcal{H}}_{\mathrm{B}}
                     + \hat{\mathcal{H}}_{\mathrm{int}},
\label{eq:Heff}
\end{equation}
where $R$ is the distance between the atom and the center-of-mass of the molecule, $\mu$ and $\hat{J}$ are the reduced mass and the total angular momentum of the collision complex, $\hat{N}$ is the rotational angular momentum of the molecule, and $\hat{S}_{\mathrm{A}}$ and $\hat{S}_{\mathrm{B}}$ are the atomic and molecular spins. The atomic Hamiltonian $\hat{\mathcal{H}}_{\mathrm{A}}  = {{g}}_e\mu_\mathrm{B} \hat{S}_{{\rm A},Z} B$, where ${g}_e$ is  the electron $\textrm{g}$-factor, $\mu_\mathrm{B}$ is the Bohr magneton, $\hat{S}_{{\rm A},Z}$ is the projection of $\hat{S}_{\rm A}$ onto the magnetic field axis and $B$ is the magnitude of the external magnetic field. 
The molecular Hamiltonian $\hat{\mathcal{H}}_{\mathrm{B}} = B_e \hat{N}^2 + \gamma_\mathrm{SR} \hat{N} \cdot \hat{S}_{\rm B} + {g}_e\mu_\mathrm{B} \hat{S}_{{\rm B},Z} B$, where $B_e$ and  $\gamma_\mathrm{SR}$ are the rotational and spin-rotation constants. In Eq. (\ref{eq:Heff}), $\hat{\mathcal{H}}_{\mathrm{int}}=\lambda\hat{{V}}+\hat{{V}}_\mathrm{dd}$ is the atom-molecule interaction, where $\hat{V}$ is the scaled atom-molecule PES discussed above and $\hat{{V}}_\mathrm{dd}$ is the magnetic dipole-dipole interaction \cite{Morita_PRA17}. 
The wave function of the collision complex is expanded in a set of basis functions \cite{Tscherbul_JPC10,Tscherbul_PRA11,Morita_RbSrF}
$\left|JM\Omega\rangle|NK_N\rangle|S_{\mathrm{A}}\Sigma_{\mathrm{A}}\rangle|S_{\mathrm{B}}\Sigma_{\mathrm{B}}\right\rangle,$
where $\Omega$, $K_N$, $\Sigma_{\mathrm{A}}$ and $\Sigma_{\mathrm{B}}$ are the projections of $J$, $N$, $S_{\mathrm{A}}$ and $S_{\mathrm{B}}$ onto the BF $z$ axis, leading to a system of CC equations, which is solved numerically as described in our previous work \cite{Tscherbul_JPC10, Tscherbul_PRA11, Morita_PRA17, Morita_RbSrF,SM}.
The size of the basis set is determined by the cutoff parameters $J_\mathrm{max}$ and $N_\mathrm{max}$ which give the maximum values of $J$ and $N$. Using $J_\mathrm{max}=3$ gives converged results over the collision energy range studied here ($E_C=10^{-6}-10^{-2}$~cm$^{-1}$).   
We consider ultracold collisions of rotationally ground-state molecules  in their maximally stretched low-field-seeking Zeeman states with spin-polarized alkali-metal atoms.

\onecolumngrid

\begin{figure}[b]
\includegraphics[width=\textwidth]{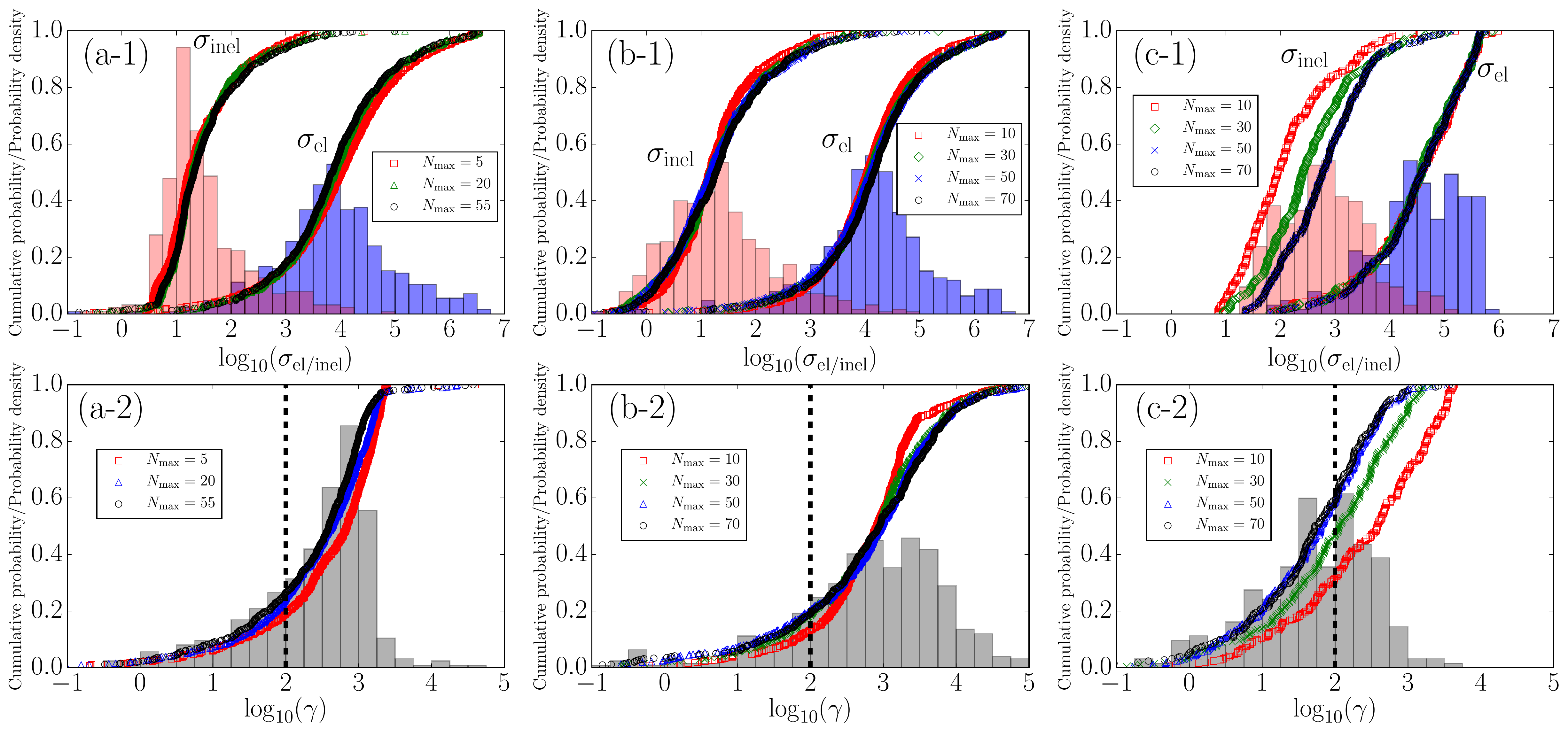}
\caption{$N_\text{max}$ dependence of the cumulative probability distributions of elastic $\sigma_\text{el}$ and inelastic $\sigma_\text{inel}$ cross sections in \AA$^2$ (upper panels) and the ratio of elastic-to-inelastic cross sections $\gamma$ (lower panels) for Li-CaH (a), Li-SrOH (b) and Rb-SrF (c) at the collision energy of $10^{-5}$ cm$^{-1}$ with the external magnetic field of 100 G. Normalized histograms show the distributions of the observables obtained with the largest $N_\text{max}$ values in each panels. At the limit of the infinity of the samples size, the histograms would converge to the probability densities.}
\label{figure2}
\end{figure}

\twocolumngrid

 \Cref{figure1} shows the cross sections for elastic scattering  and inelastic spin relaxation in Li-CaH collisions as functions of the scaling parameter $\lambda$. The numerous scattering resonances lead to a large variation  of the cross sections over a narrow interval of $\lambda$.
We observe a dramatic decrease of the resonance density  with decreasing the cutoff parameter $N_\mathrm{max}$ due to a decrease in the number of bound states of the collision complex.

 
In \cref{figure2}, we show the CPDs of the scattering cross sections for Li-CaH, Li-SrOH and Rb-SrF obtained in CC calculations with different basis sets.
Remarkably, the CPDs  converge  quickly with respect to basis set size, as shown for Li-CaH in \cref{figure2}(a-1) and (a-2).
This demonstrates that we can obtain accurate CPDs using severely restricted CC basis sets that would be to small to produce fully converged scattering cross sections. This opens up the prospect of  computing accurate CPDs for molecular systems with many degrees of freedom, for which fully converged CC calculations are prohibitively difficult.

From \cref{figure2}(a-2), the probability for the elastic-to-inelastic ratio  $\gamma$ to fall below 100 is  $\sim$0.2, giving the success probability of sympathetic cooling of 80\%. 
The success probability derived from the CPD thus provides a useful parameter for screening the suitability of atom-molecule systems for  sympathetic cooling based on their collisional properties. 
The rapid convergence of CPDs with respect to $N_\text{max}$ is an important advantage of the CPDs over the individual $\gamma$ values used previously for this purpose \cite{Rev_Carr08,Tscherbul_PRA11,Hutson_PRL13}.

Figures~\ref{figure2}(b-1) and (b-2) show the results for a heavier collision system Li-SrOH, which requires much larger rotational basis sets than Li-CaH ($N_\text{max}=115$)  to obtain converged results  due to the  small rotational constant of SrOH \cite{SM,Morita_PRA17}. 
Despite this, we observe rapid convergence of the CPDs with respect to $N_\text{max}$,  making it possible to use   basis sets with $N_\text{max}=30$ for quantitatively accurate calculations. 

The physical reason behind the rapid convergence of CPDs with respect to $N_\text{max}$ is that the ensemble averaging procedure washes out the intricate details of the resonance  structure shown in Fig.~\ref{figure1}, provided that the $\lambda$ interval  contains enough resonances.
In addition, the CPDs do not contain information about the correlation between the values of $\lambda$ and the those of the scattering observables. For example, if we consider a periodic function of $\lambda$,  the CPD of the periodic function is invariant with respect to the change of the period of oscillations as long as the function oscillates many times in the chosen interval of $\lambda$. Thus, while we observe a significant difference in the resonance density as a function of $\lambda$ between \cref{figure1} (a) and (b), such differences tend to have no effect on the CPDs.

\begin{figure}[t]
\includegraphics[width=\linewidth]{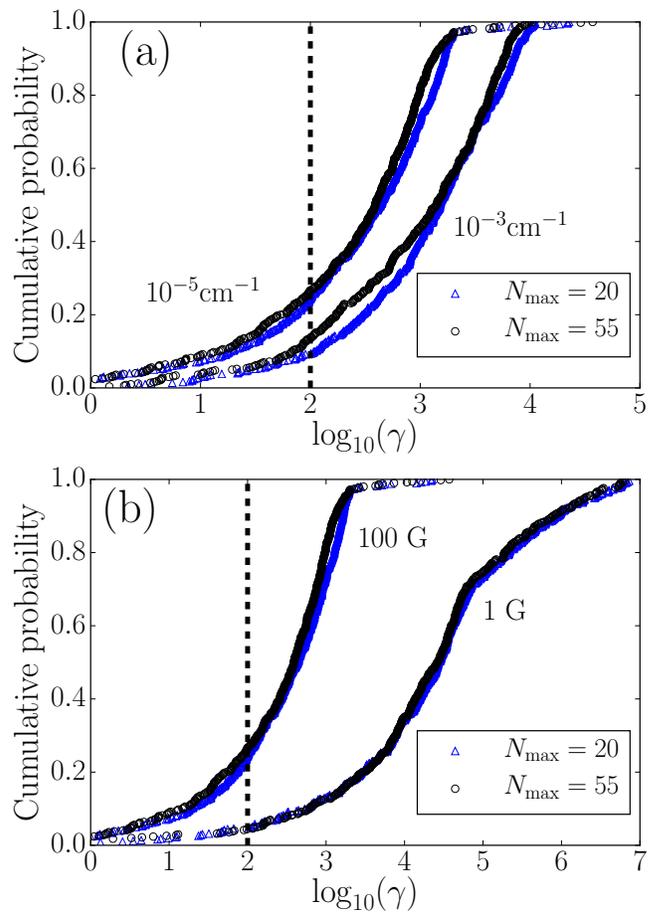}
\caption{Comparison of the cumulative probability distributions of the elastic-to-inelastic ratio for Li-CaH (a)
between $E=10^{-5}$ cm$^{-1}$ and $E=10^{-3}$ cm$^{-1}$ with the external magnetic field of 100 G and (b) between $B=100$ G vs $B=1$ G at the collision energy of $10^{-5}$ cm$^{-1}$.}  
\label{figure3}
\end{figure}

Figure~\ref{figure2}(c-1) shows the CPDs calculated for the heaviest collision system studied here (Rb-SrF), which also exhibits the slowest basis set convergence, with $N_\text{max}=125$ required to produce accurate results \cite{Morita_RbSrF}.
The CPDs of the inelastic cross sections (but not the elastic ones!) tend toward higher values of $\sigma_\text{inel}$  with increasing $N_\text{max}$, saturating only at $N_\text{max}=50$, which is still significantly lower than required for the fully converged CC calculation \cite{Morita_RbSrF}.  We attribute the slow convergence rate to the broad  shape resonances that occur in the inelastic cross sections \cite{Morita_RbSrF}, affecting their background values over a large range of magnetic fields.  This also indicates that the couplings with very highly excited rotational channels $N > 30$ play an important role for the background value of the inelastic cross section due to the strong anisotropy of the Rb-SrF interaction.  

To explore the variation of the CPDs with the collision energy and magnetic field, we compare in \Cref{figure3}(a) the CPDs of the elastic-to-inelastic ratio $\gamma$ for Li-CaH in the $s$-wave regime ($E_C=10^{-5}$ cm$^{-1}$) vs the multiple partial-wave regime  ($E_C=10^{-3}$ cm$^{-1}$). 
\Cref{figure3}(b) shows a similar plot for two different values of the magnetic field.  
We observe that the CPDs at different collision energies and magnetic fields are  statistically distinguishable  from each other, even within the limits imposed by numerical convergence, making it possible to  reliably predict the variation of the success probability with collision energy and field. 
The results shown in Fig.~\ref{figure3} also suggest that  the rapid convergence of CPDs with respect to $N_\text{max}$ holds regardless of the  collision energy and magnetic field.  
We can therefore systematically examine the success probability of sympathetic cooling as a function of the collision energy and magnetic field using modest basis sets with $N_\mathrm{max}=20$ for Li-CaH. 

\begin{figure}[t]
\begin{center}
\includegraphics[width=\linewidth]{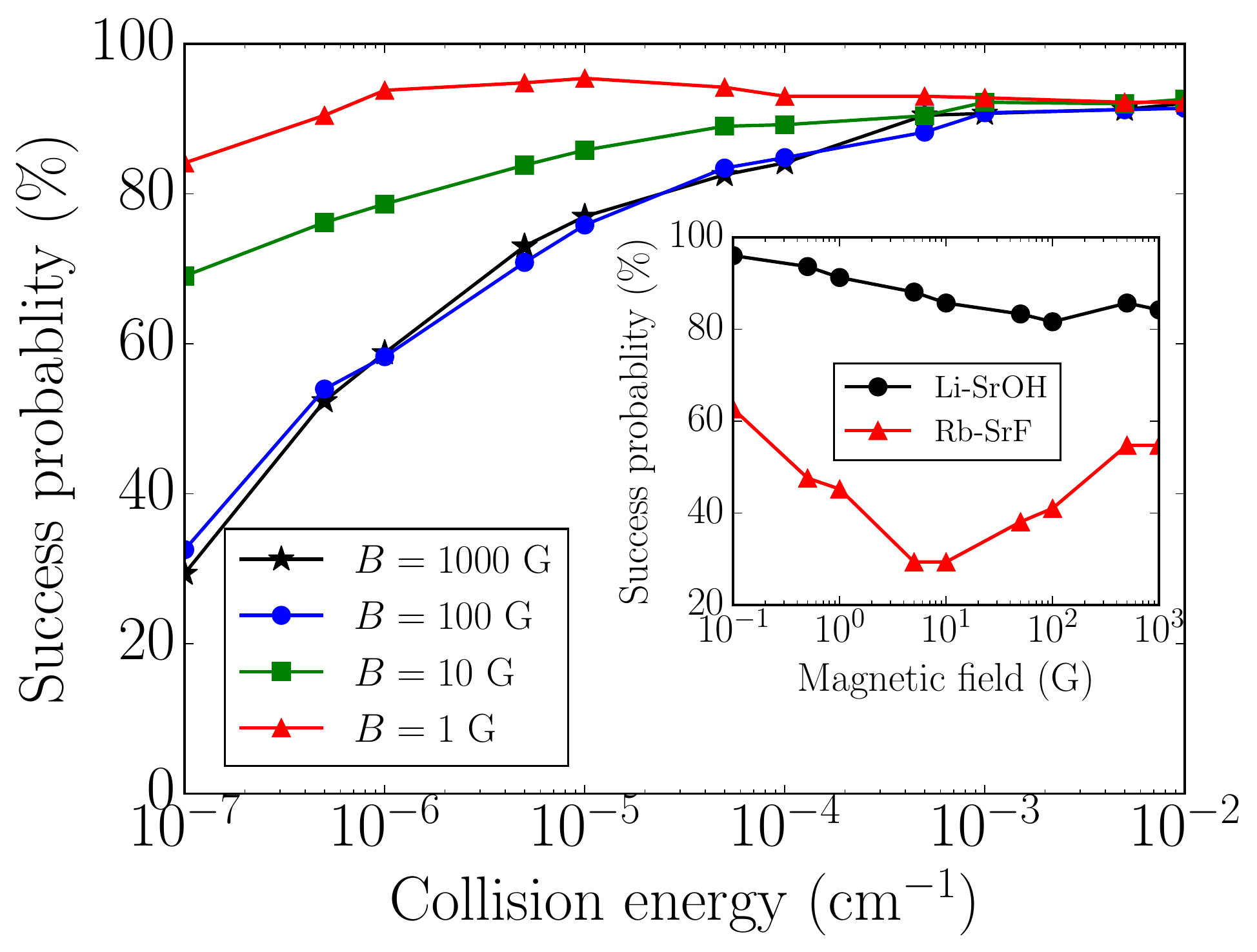}
\end{center}
\caption{ (a) Success probability of sympathetic cooling of Li-CaH as a function of collision energy for the external magnetic field of 1000G (black stars) 100 G (blue circles), 10 G (green squares) and 1 G (red triangles). $N_\text{max}=20$. (b) Success probability of sympathetic cooling of Li-SrOH (black circles) and Rb-SrF (red triangles) as a function of the magnitude of external magnetic field at the collision energy of $10^{-5}$ cm$^{-1}$.  $N_\text{max}=30$ for Li-SrOH and  $N_\text{max}=50$ for Rb-SrF. 
}
\label{figure4}
\end{figure}

Figure~\ref{figure4} shows the success probability of Li-CaH sympathetic cooling as a function of collision energy. At  $E_C\sim 1$~mK, the probability is high regardless of the magnitude of the magnetic field. At lower collision energies, the success probability decreases by a factor of 3-5 with increasing field. A significant magnetic field dependence of the success probability for Rb-SrF shown in the inset of Figure~\ref{figure4}  suggests that varying the field could be used to optimize sympathetic cooling  of $^2\Sigma$ molecules with alkali-metal atoms in a magnetic trap \cite{Morita_RbSrF}

In summary, we have proposed a new statistical approach to account for the effect of PES uncertainties on cold collision observables.  The approach is based on the CPDs computed by averaging the results of quantum scattering calculations over an ensemble of interaction PES produced by the potential scaling method. 
The remarkably fast basis set convergence of the CPDs makes our approach applicable to a wide range of molecular collision systems of current experimental interest, such as polyatomic molecules with low-frequency vibrational modes \cite{Kozyryev:17,Momose_PRL17} and ultracold chemical reactions \cite{Rev_Bala16}.  It would be interesting to further explore the properties of CPDs in connection with the statistical model of insertion chemical reactions \cite{Miller:70}  \cite{Croft:17}, and with the theory of quantum chaotic scattering \cite{Frisch:14,Croft:14,Mayle:13}. 

\begin{acknowledgments}
We are grateful to Balakrishnan Naduvalath for stimulating discussions. This work was supported by NSF grant No. PHY-1607610.
\end{acknowledgments}



\bibliography{cite.bib}

\begin{thebibliography}{35}%
\makeatletter
\providecommand \@ifxundefined [1]{%
 \@ifx{#1\undefined}
}%
\providecommand \@ifnum [1]{%
 \ifnum #1\expandafter \@firstoftwo
 \else \expandafter \@secondoftwo
 \fi
}%
\providecommand \@ifx [1]{%
 \ifx #1\expandafter \@firstoftwo
 \else \expandafter \@secondoftwo
 \fi
}%
\providecommand \natexlab [1]{#1}%
\providecommand \enquote  [1]{``#1''}%
\providecommand \bibnamefont  [1]{#1}%
\providecommand \bibfnamefont [1]{#1}%
\providecommand \citenamefont [1]{#1}%
\providecommand \href@noop [0]{\@secondoftwo}%
\providecommand \href [0]{\begingroup \@sanitize@url \@href}%
\providecommand \@href[1]{\@@startlink{#1}\@@href}%
\providecommand \@@href[1]{\endgroup#1\@@endlink}%
\providecommand \@sanitize@url [0]{\catcode `\\12\catcode `\$12\catcode
  `\&12\catcode `\#12\catcode `\^12\catcode `\_12\catcode `\%12\relax}%
\providecommand \@@startlink[1]{}%
\providecommand \@@endlink[0]{}%
\providecommand \url  [0]{\begingroup\@sanitize@url \@url }%
\providecommand \@url [1]{\endgroup\@href {#1}{\urlprefix }}%
\providecommand \urlprefix  [0]{URL }%
\providecommand \Eprint [0]{\href }%
\providecommand \doibase [0]{http://dx.doi.org/}%
\providecommand \selectlanguage [0]{\@gobble}%
\providecommand \bibinfo  [0]{\@secondoftwo}%
\providecommand \bibfield  [0]{\@secondoftwo}%
\providecommand \translation [1]{[#1]}%
\providecommand \BibitemOpen [0]{}%
\providecommand \bibitemStop [0]{}%
\providecommand \bibitemNoStop [0]{.\EOS\space}%
\providecommand \EOS [0]{\spacefactor3000\relax}%
\providecommand \BibitemShut  [1]{\csname bibitem#1\endcsname}%
\let\auto@bib@innerbib\@empty
\bibitem [{\citenamefont {Carr}\ \emph {et~al.}(2009)\citenamefont {Carr},
  \citenamefont {DeMille}, \citenamefont {Krems},\ and\ \citenamefont
  {Ye}}]{Rev_Carr08}%
  \BibitemOpen
  \bibfield  {author} {\bibinfo {author} {\bibfnamefont {L.~D.}\ \bibnamefont
  {Carr}}, \bibinfo {author} {\bibfnamefont {D.}~\bibnamefont {DeMille}},
  \bibinfo {author} {\bibfnamefont {R.~V.}\ \bibnamefont {Krems}}, \ and\
  \bibinfo {author} {\bibfnamefont {J.}~\bibnamefont {Ye}},\ }\href
  {http://stacks.iop.org/1367-2630/11/i=5/a=055049} {\bibfield  {journal}
  {\bibinfo  {journal} {New Journal of Physics}\ }\textbf {\bibinfo {volume}
  {11}},\ \bibinfo {pages} {055049} (\bibinfo {year} {2009})}\BibitemShut
  {NoStop}%
\bibitem [{\citenamefont {Bohn}\ \emph {et~al.}(2017)\citenamefont {Bohn},
  \citenamefont {Rey},\ and\ \citenamefont {Ye}}]{Rev_Bohn17}%
  \BibitemOpen
  \bibfield  {author} {\bibinfo {author} {\bibfnamefont {J.~L.}\ \bibnamefont
  {Bohn}}, \bibinfo {author} {\bibfnamefont {A.~M.}\ \bibnamefont {Rey}}, \
  and\ \bibinfo {author} {\bibfnamefont {J.}~\bibnamefont {Ye}},\ }\href
  {\doibase 10.1126/science.aam6299} {\bibfield  {journal} {\bibinfo  {journal}
  {Science}\ }\textbf {\bibinfo {volume} {357}},\ \bibinfo {pages} {1002}
  (\bibinfo {year} {2017})}\BibitemShut {NoStop}%
\bibitem [{\citenamefont {Moses}\ \emph {et~al.}(2016)\citenamefont {Moses},
  \citenamefont {Covey}, \citenamefont {Miecnikowski}, \citenamefont {Jin},\
  and\ \citenamefont {Ye}}]{Moses:17}%
  \BibitemOpen
  \bibfield  {author} {\bibinfo {author} {\bibfnamefont {S.~A.}\ \bibnamefont
  {Moses}}, \bibinfo {author} {\bibfnamefont {J.~P.}\ \bibnamefont {Covey}},
  \bibinfo {author} {\bibfnamefont {M.~T.}\ \bibnamefont {Miecnikowski}},
  \bibinfo {author} {\bibfnamefont {D.~S.}\ \bibnamefont {Jin}}, \ and\
  \bibinfo {author} {\bibfnamefont {J.}~\bibnamefont {Ye}},\ }\href@noop {}
  {\bibfield  {journal} {\bibinfo  {journal} {Nature Physics}\ }\textbf
  {\bibinfo {volume} {13}},\ \bibinfo {pages} {13} (\bibinfo {year}
  {2016})}\BibitemShut {NoStop}%
\bibitem [{\citenamefont {Krems}(2018)}]{Krems:18}%
  \BibitemOpen
  \bibfield  {author} {\bibinfo {author} {\bibfnamefont {R.~V.}\ \bibnamefont
  {Krems}},\ }\enquote {\bibinfo {title} {{\it Molecules in electromagnetic
  fields: from ultracold physics to controlled chemistry}},}\ \ (\bibinfo
  {publisher} {Wiley},\ \bibinfo {year} {2018})\BibitemShut {NoStop}%
\bibitem [{\citenamefont {Krems}(2008)}]{Rev_Krems08}%
  \BibitemOpen
  \bibfield  {author} {\bibinfo {author} {\bibfnamefont {R.~V.}\ \bibnamefont
  {Krems}},\ }\href {\doibase 10.1039/B802322K} {\bibfield  {journal} {\bibinfo
   {journal} {Phys. Chem. Chem. Phys.}\ }\textbf {\bibinfo {volume} {10}},\
  \bibinfo {pages} {4079} (\bibinfo {year} {2008})}\BibitemShut {NoStop}%
\bibitem [{\citenamefont {DeMille}(2002)}]{DeMille:02}%
  \BibitemOpen
  \bibfield  {author} {\bibinfo {author} {\bibfnamefont {D.}~\bibnamefont
  {DeMille}},\ }\href {\doibase 10.1103/PhysRevLett.88.067901} {\bibfield
  {journal} {\bibinfo  {journal} {Phys. Rev. Lett.}\ }\textbf {\bibinfo
  {volume} {88}},\ \bibinfo {pages} {067901} (\bibinfo {year}
  {2002})}\BibitemShut {NoStop}%
\bibitem [{\citenamefont {Safronova}\ \emph {et~al.}(2018)\citenamefont
  {Safronova}, \citenamefont {Budker}, \citenamefont {DeMille}, \citenamefont
  {Kimball}, \citenamefont {Derevianko},\ and\ \citenamefont
  {Clark}}]{Rev_Safronova18}%
  \BibitemOpen
  \bibfield  {author} {\bibinfo {author} {\bibfnamefont {M.~S.}\ \bibnamefont
  {Safronova}}, \bibinfo {author} {\bibfnamefont {D.}~\bibnamefont {Budker}},
  \bibinfo {author} {\bibfnamefont {D.}~\bibnamefont {DeMille}}, \bibinfo
  {author} {\bibfnamefont {D.~F.~J.}\ \bibnamefont {Kimball}}, \bibinfo
  {author} {\bibfnamefont {A.}~\bibnamefont {Derevianko}}, \ and\ \bibinfo
  {author} {\bibfnamefont {C.~W.}\ \bibnamefont {Clark}},\ }\href {\doibase
  10.1103/RevModPhys.90.025008} {\bibfield  {journal} {\bibinfo  {journal}
  {Rev. Mod. Phys.}\ }\textbf {\bibinfo {volume} {90}},\ \bibinfo {pages}
  {025008} (\bibinfo {year} {2018})}\BibitemShut {NoStop}%
\bibitem [{\citenamefont {McCarron}\ \emph {et~al.}(2018)\citenamefont
  {McCarron}, \citenamefont {Steinecker}, \citenamefont {Zhu},\ and\
  \citenamefont {DeMille}}]{DeMIlle_PRL18}%
  \BibitemOpen
  \bibfield  {author} {\bibinfo {author} {\bibfnamefont {D.~J.}\ \bibnamefont
  {McCarron}}, \bibinfo {author} {\bibfnamefont {M.~H.}\ \bibnamefont
  {Steinecker}}, \bibinfo {author} {\bibfnamefont {Y.}~\bibnamefont {Zhu}}, \
  and\ \bibinfo {author} {\bibfnamefont {D.}~\bibnamefont {DeMille}},\ }\href
  {\doibase 10.1103/PhysRevLett.121.013202} {\bibfield  {journal} {\bibinfo
  {journal} {Phys. Rev. Lett.}\ }\textbf {\bibinfo {volume} {121}},\ \bibinfo
  {pages} {013202} (\bibinfo {year} {2018})}\BibitemShut {NoStop}%
\bibitem [{\citenamefont {Anderegg}\ \emph {et~al.}(2018)\citenamefont
  {Anderegg}, \citenamefont {Augenbraun}, \citenamefont {Bao}, \citenamefont
  {Burchesky}, \citenamefont {Cheuk}, \citenamefont {Ketterle},\ and\
  \citenamefont {Doyle}}]{Anderegg_Nature18}%
  \BibitemOpen
  \bibfield  {author} {\bibinfo {author} {\bibfnamefont {L.}~\bibnamefont
  {Anderegg}}, \bibinfo {author} {\bibfnamefont {B.~L.}\ \bibnamefont
  {Augenbraun}}, \bibinfo {author} {\bibfnamefont {Y.}~\bibnamefont {Bao}},
  \bibinfo {author} {\bibfnamefont {S.}~\bibnamefont {Burchesky}}, \bibinfo
  {author} {\bibfnamefont {L.~W.}\ \bibnamefont {Cheuk}}, \bibinfo {author}
  {\bibfnamefont {W.}~\bibnamefont {Ketterle}}, \ and\ \bibinfo {author}
  {\bibfnamefont {J.~M.}\ \bibnamefont {Doyle}},\ }\href@noop {} {\bibfield
  {journal} {\bibinfo  {journal} {Nature Physics}\ }\textbf {\bibinfo {volume}
  {14}},\ \bibinfo {pages} {890} (\bibinfo {year} {2018})}\BibitemShut
  {NoStop}%
\bibitem [{\citenamefont {Klein}\ \emph {et~al.}(2016)\citenamefont {Klein},
  \citenamefont {Shagam}, \citenamefont {Skomorowski}, \citenamefont
  {{\.Z}uchowski}, \citenamefont {Pawlak}, \citenamefont {Janssen},
  \citenamefont {Moiseyev}, \citenamefont {van~de Meerakker}, \citenamefont
  {van~der Avoird}, \citenamefont {Koch},\ and\ \citenamefont
  {Narevicius}}]{Klein:16}%
  \BibitemOpen
  \bibfield  {author} {\bibinfo {author} {\bibfnamefont {A.}~\bibnamefont
  {Klein}}, \bibinfo {author} {\bibfnamefont {Y.}~\bibnamefont {Shagam}},
  \bibinfo {author} {\bibfnamefont {W.}~\bibnamefont {Skomorowski}}, \bibinfo
  {author} {\bibfnamefont {P.~S.}\ \bibnamefont {{\.Z}uchowski}}, \bibinfo
  {author} {\bibfnamefont {M.}~\bibnamefont {Pawlak}}, \bibinfo {author}
  {\bibfnamefont {L.~M.~C.}\ \bibnamefont {Janssen}}, \bibinfo {author}
  {\bibfnamefont {N.}~\bibnamefont {Moiseyev}}, \bibinfo {author}
  {\bibfnamefont {S.~Y.~T.}\ \bibnamefont {van~de Meerakker}}, \bibinfo
  {author} {\bibfnamefont {A.}~\bibnamefont {van~der Avoird}}, \bibinfo
  {author} {\bibfnamefont {C.~P.}\ \bibnamefont {Koch}}, \ and\ \bibinfo
  {author} {\bibfnamefont {E.}~\bibnamefont {Narevicius}},\ }\href@noop {}
  {\bibfield  {journal} {\bibinfo  {journal} {Nature Physics}\ }\textbf
  {\bibinfo {volume} {13}},\ \bibinfo {pages} {35} (\bibinfo {year}
  {2016})}\BibitemShut {NoStop}%
\bibitem [{\citenamefont {Perreault}\ \emph {et~al.}(2017)\citenamefont
  {Perreault}, \citenamefont {Mukherjee},\ and\ \citenamefont
  {Zare}}]{Perreault:17}%
  \BibitemOpen
  \bibfield  {author} {\bibinfo {author} {\bibfnamefont {W.~E.}\ \bibnamefont
  {Perreault}}, \bibinfo {author} {\bibfnamefont {N.}~\bibnamefont
  {Mukherjee}}, \ and\ \bibinfo {author} {\bibfnamefont {R.~N.}\ \bibnamefont
  {Zare}},\ }\href@noop {} {\bibfield  {journal} {\bibinfo  {journal}
  {Science}\ }\textbf {\bibinfo {volume} {358}},\ \bibinfo {pages} {356}
  (\bibinfo {year} {2017})}\BibitemShut {NoStop}%
\bibitem [{\citenamefont {Ni}\ \emph {et~al.}(2010)\citenamefont {Ni},
  \citenamefont {Ospelkaus}, \citenamefont {Wang}, \citenamefont
  {Qu{\'e}m{\'e}ner}, \citenamefont {Neyenhuis}, \citenamefont {de~Miranda},
  \citenamefont {Bohn}, \citenamefont {Ye},\ and\ \citenamefont
  {Jin}}]{Ni_NATURE10}%
  \BibitemOpen
  \bibfield  {author} {\bibinfo {author} {\bibfnamefont {K.~K.}\ \bibnamefont
  {Ni}}, \bibinfo {author} {\bibfnamefont {S.}~\bibnamefont {Ospelkaus}},
  \bibinfo {author} {\bibfnamefont {D.}~\bibnamefont {Wang}}, \bibinfo {author}
  {\bibfnamefont {G.}~\bibnamefont {Qu{\'e}m{\'e}ner}}, \bibinfo {author}
  {\bibfnamefont {B.}~\bibnamefont {Neyenhuis}}, \bibinfo {author}
  {\bibfnamefont {M.~H.~G.}\ \bibnamefont {de~Miranda}}, \bibinfo {author}
  {\bibfnamefont {J.~L.}\ \bibnamefont {Bohn}}, \bibinfo {author}
  {\bibfnamefont {J.}~\bibnamefont {Ye}}, \ and\ \bibinfo {author}
  {\bibfnamefont {D.~S.}\ \bibnamefont {Jin}},\ }\href
  {http://dx.doi.org/10.1038/nature08953} {\bibfield  {journal} {\bibinfo
  {journal} {Nature}\ }\textbf {\bibinfo {volume} {464}},\ \bibinfo {pages}
  {1324} (\bibinfo {year} {2010})}\BibitemShut {NoStop}%
\bibitem [{\citenamefont {Balakrishnan}(2016)}]{Rev_Bala16}%
  \BibitemOpen
  \bibfield  {author} {\bibinfo {author} {\bibfnamefont {N.}~\bibnamefont
  {Balakrishnan}},\ }\href {\doibase 10.1063/1.4964096} {\bibfield  {journal}
  {\bibinfo  {journal} {J. Chem. Phys.}\ }\textbf {\bibinfo {volume} {145}},\
  \bibinfo {pages} {150901} (\bibinfo {year} {2016})}\BibitemShut {NoStop}%
\bibitem [{\citenamefont {de~Miranda}\ \emph {et~al.}(2011)\citenamefont
  {de~Miranda}, \citenamefont {Chotia}, \citenamefont {Neyenhuis},
  \citenamefont {Wang}, \citenamefont {Qu{\'e}m{\'e}ner}, \citenamefont
  {Ospelkaus}, \citenamefont {Bohn}, \citenamefont {Ye},\ and\ \citenamefont
  {Jin}}]{Miranda:11}%
  \BibitemOpen
  \bibfield  {author} {\bibinfo {author} {\bibfnamefont {M.~H.~G.}\
  \bibnamefont {de~Miranda}}, \bibinfo {author} {\bibfnamefont
  {A.}~\bibnamefont {Chotia}}, \bibinfo {author} {\bibfnamefont
  {B.}~\bibnamefont {Neyenhuis}}, \bibinfo {author} {\bibfnamefont
  {D.}~\bibnamefont {Wang}}, \bibinfo {author} {\bibfnamefont {G.}~\bibnamefont
  {Qu{\'e}m{\'e}ner}}, \bibinfo {author} {\bibfnamefont {S.}~\bibnamefont
  {Ospelkaus}}, \bibinfo {author} {\bibfnamefont {J.~L.}\ \bibnamefont {Bohn}},
  \bibinfo {author} {\bibfnamefont {J.}~\bibnamefont {Ye}}, \ and\ \bibinfo
  {author} {\bibfnamefont {D.~S.}\ \bibnamefont {Jin}},\ }\href@noop {}
  {\bibfield  {journal} {\bibinfo  {journal} {Nature Physics}\ }\textbf
  {\bibinfo {volume} {7}},\ \bibinfo {pages} {502} (\bibinfo {year}
  {2011})}\BibitemShut {NoStop}%
\bibitem [{\citenamefont {Yan}\ \emph {et~al.}(2013)\citenamefont {Yan},
  \citenamefont {Moses}, \citenamefont {Gadway}, \citenamefont {Covey},
  \citenamefont {Hazzard}, \citenamefont {Rey}, \citenamefont {Jin},\ and\
  \citenamefont {Ye}}]{Yan:13}%
  \BibitemOpen
  \bibfield  {author} {\bibinfo {author} {\bibfnamefont {B.}~\bibnamefont
  {Yan}}, \bibinfo {author} {\bibfnamefont {S.~A.}\ \bibnamefont {Moses}},
  \bibinfo {author} {\bibfnamefont {B.}~\bibnamefont {Gadway}}, \bibinfo
  {author} {\bibfnamefont {J.~P.}\ \bibnamefont {Covey}}, \bibinfo {author}
  {\bibfnamefont {K.~R.~A.}\ \bibnamefont {Hazzard}}, \bibinfo {author}
  {\bibfnamefont {A.~M.}\ \bibnamefont {Rey}}, \bibinfo {author} {\bibfnamefont
  {D.~S.}\ \bibnamefont {Jin}}, \ and\ \bibinfo {author} {\bibfnamefont
  {J.}~\bibnamefont {Ye}},\ }\href@noop {} {\bibfield  {journal} {\bibinfo
  {journal} {Nature (London)}\ }\textbf {\bibinfo {volume} {501}},\ \bibinfo
  {pages} {521} (\bibinfo {year} {2013})}\BibitemShut {NoStop}%
\bibitem [{\citenamefont {Andreev}\ \emph {et~al.}(2018)\citenamefont
  {Andreev}, \citenamefont {Ang}, \citenamefont {DeMille}, \citenamefont
  {Doyle}, \citenamefont {Gabrielse}, \citenamefont {Haefner}, \citenamefont
  {Hutzler}, \citenamefont {Lasner}, \citenamefont {Meisenhelder},
  \citenamefont {O'Leary}, \citenamefont {Panda}, \citenamefont {West},
  \citenamefont {West}, \citenamefont {Wu},\ and\ \citenamefont
  {Collaboration}}]{Andreev:18}%
  \BibitemOpen
  \bibfield  {author} {\bibinfo {author} {\bibfnamefont {V.}~\bibnamefont
  {Andreev}}, \bibinfo {author} {\bibfnamefont {D.~G.}\ \bibnamefont {Ang}},
  \bibinfo {author} {\bibfnamefont {D.}~\bibnamefont {DeMille}}, \bibinfo
  {author} {\bibfnamefont {J.~M.}\ \bibnamefont {Doyle}}, \bibinfo {author}
  {\bibfnamefont {G.}~\bibnamefont {Gabrielse}}, \bibinfo {author}
  {\bibfnamefont {J.}~\bibnamefont {Haefner}}, \bibinfo {author} {\bibfnamefont
  {N.~R.}\ \bibnamefont {Hutzler}}, \bibinfo {author} {\bibfnamefont
  {Z.}~\bibnamefont {Lasner}}, \bibinfo {author} {\bibfnamefont
  {C.}~\bibnamefont {Meisenhelder}}, \bibinfo {author} {\bibfnamefont {B.~R.}\
  \bibnamefont {O'Leary}}, \bibinfo {author} {\bibfnamefont {C.~D.}\
  \bibnamefont {Panda}}, \bibinfo {author} {\bibfnamefont {A.~D.}\ \bibnamefont
  {West}}, \bibinfo {author} {\bibfnamefont {E.~P.}\ \bibnamefont {West}},
  \bibinfo {author} {\bibfnamefont {X.}~\bibnamefont {Wu}}, \ and\ \bibinfo
  {author} {\bibfnamefont {{\it ACME}.}~\bibnamefont {Collaboration}},\
  }\href@noop {} {\bibfield  {journal} {\bibinfo  {journal} {Nature}\ }\textbf
  {\bibinfo {volume} {562}},\ \bibinfo {pages} {355} (\bibinfo {year}
  {2018})}\BibitemShut {NoStop}%
\bibitem [{\citenamefont {Lemeshko}\ \emph {et~al.}(2013)\citenamefont
  {Lemeshko}, \citenamefont {Krems}, \citenamefont {Doyle},\ and\ \citenamefont
  {Kais}}]{Lemeshko:13}%
  \BibitemOpen
  \bibfield  {author} {\bibinfo {author} {\bibfnamefont {M.}~\bibnamefont
  {Lemeshko}}, \bibinfo {author} {\bibfnamefont {R.~V.}\ \bibnamefont {Krems}},
  \bibinfo {author} {\bibfnamefont {J.~M.}\ \bibnamefont {Doyle}}, \ and\
  \bibinfo {author} {\bibfnamefont {S.}~\bibnamefont {Kais}},\ }\href@noop {}
  {\bibfield  {journal} {\bibinfo  {journal} {Mol. Phys}\ }\textbf {\bibinfo
  {volume} {111}},\ \bibinfo {pages} {1648} (\bibinfo {year}
  {2013})}\BibitemShut {NoStop}%
\bibitem [{\citenamefont {Wallis}\ and\ \citenamefont
  {Hutson}(2009)}]{Hutson_PRL09}%
  \BibitemOpen
  \bibfield  {author} {\bibinfo {author} {\bibfnamefont {A.~O.~G.}\
  \bibnamefont {Wallis}}\ and\ \bibinfo {author} {\bibfnamefont {J.~M.}\
  \bibnamefont {Hutson}},\ }\href {\doibase 10.1103/PhysRevLett.103.183201}
  {\bibfield  {journal} {\bibinfo  {journal} {Phys. Rev. Lett.}\ }\textbf
  {\bibinfo {volume} {103}},\ \bibinfo {pages} {183201} (\bibinfo {year}
  {2009})}\BibitemShut {NoStop}%
\bibitem [{\citenamefont {Wallis}\ \emph {et~al.}(2011)\citenamefont {Wallis},
  \citenamefont {Longdon}, \citenamefont {{\.{Z}}uchowski},\ and\ \citenamefont
  {Hutson}}]{Hutson_EPJD11}%
  \BibitemOpen
  \bibfield  {author} {\bibinfo {author} {\bibfnamefont {A.~O.~G.}\
  \bibnamefont {Wallis}}, \bibinfo {author} {\bibfnamefont {E.~J.~J.}\
  \bibnamefont {Longdon}}, \bibinfo {author} {\bibfnamefont {P.~S.}\
  \bibnamefont {{\.{Z}}uchowski}}, \ and\ \bibinfo {author} {\bibfnamefont
  {J.~M.}\ \bibnamefont {Hutson}},\ }\href {\doibase
  10.1140/epjd/e2011-20025-4} {\bibfield  {journal} {\bibinfo  {journal} {The
  European Physical Journal D}\ }\textbf {\bibinfo {volume} {65}},\ \bibinfo
  {pages} {151} (\bibinfo {year} {2011})}\BibitemShut {NoStop}%
\bibitem [{\citenamefont {{\.Z}uchowski}\ and\ \citenamefont
  {Hutson}(2011)}]{Hutson_PCCP11}%
  \BibitemOpen
  \bibfield  {author} {\bibinfo {author} {\bibfnamefont {P.~S.}\ \bibnamefont
  {{\.Z}uchowski}}\ and\ \bibinfo {author} {\bibfnamefont {J.~M.}\ \bibnamefont
  {Hutson}},\ }\href {\doibase 10.1039/C0CP01447H} {\bibfield  {journal}
  {\bibinfo  {journal} {Phys. Chem. Chem. Phys.}\ }\textbf {\bibinfo {volume}
  {13}},\ \bibinfo {pages} {3669} (\bibinfo {year} {2011})}\BibitemShut
  {NoStop}%
\bibitem [{\citenamefont {Cui}\ and\ \citenamefont
  {Krems}(2013)}]{Krems_PRA13}%
  \BibitemOpen
  \bibfield  {author} {\bibinfo {author} {\bibfnamefont {J.}~\bibnamefont
  {Cui}}\ and\ \bibinfo {author} {\bibfnamefont {R.~V.}\ \bibnamefont
  {Krems}},\ }\href {\doibase 10.1103/PhysRevA.88.042705} {\bibfield  {journal}
  {\bibinfo  {journal} {Phys. Rev. A}\ }\textbf {\bibinfo {volume} {88}},\
  \bibinfo {pages} {042705} (\bibinfo {year} {2013})}\BibitemShut {NoStop}%
\bibitem [{\citenamefont {Suleimanov}\ and\ \citenamefont
  {Tscherbul}(2016)}]{Suleimanov_JPB16}%
  \BibitemOpen
  \bibfield  {author} {\bibinfo {author} {\bibfnamefont {Y.~V.}\ \bibnamefont
  {Suleimanov}}\ and\ \bibinfo {author} {\bibfnamefont {T.~V.}\ \bibnamefont
  {Tscherbul}},\ }\href {http://stacks.iop.org/0953-4075/49/i=20/a=204002}
  {\bibfield  {journal} {\bibinfo  {journal} {Journal of Physics B: Atomic,
  Molecular and Optical Physics}\ }\textbf {\bibinfo {volume} {49}},\ \bibinfo
  {pages} {204002} (\bibinfo {year} {2016})}\BibitemShut {NoStop}%
\bibitem [{\citenamefont {Tscherbul}\ \emph {et~al.}(2011)\citenamefont
  {Tscherbul}, \citenamefont {K\l{}os},\ and\ \citenamefont
  {Buchachenko}}]{Tscherbul_PRA11}%
  \BibitemOpen
  \bibfield  {author} {\bibinfo {author} {\bibfnamefont {T.~V.}\ \bibnamefont
  {Tscherbul}}, \bibinfo {author} {\bibfnamefont {J.}~\bibnamefont {K\l{}os}},
  \ and\ \bibinfo {author} {\bibfnamefont {A.~A.}\ \bibnamefont
  {Buchachenko}},\ }\href {\doibase 10.1103/PhysRevA.84.040701} {\bibfield
  {journal} {\bibinfo  {journal} {Phys. Rev. A}\ }\textbf {\bibinfo {volume}
  {84}},\ \bibinfo {pages} {040701} (\bibinfo {year} {2011})}\BibitemShut
  {NoStop}%
\bibitem [{\citenamefont {Morita}\ \emph {et~al.}(2018)\citenamefont {Morita},
  \citenamefont {Kosicki}, \citenamefont {\ifmmode~\dot{Z}\else
  \.{Z}\fi{}uchowski},\ and\ \citenamefont {Tscherbul}}]{Morita_RbSrF}%
  \BibitemOpen
  \bibfield  {author} {\bibinfo {author} {\bibfnamefont {M.}~\bibnamefont
  {Morita}}, \bibinfo {author} {\bibfnamefont {M.~B.}\ \bibnamefont {Kosicki}},
  \bibinfo {author} {\bibfnamefont {P.~S.}\ \bibnamefont {\ifmmode~\dot{Z}\else
  \.{Z}\fi{}uchowski}}, \ and\ \bibinfo {author} {\bibfnamefont {T.~V.}\
  \bibnamefont {Tscherbul}},\ }\href {\doibase 10.1103/PhysRevA.98.042702}
  {\bibfield  {journal} {\bibinfo  {journal} {Phys. Rev. A}\ }\textbf {\bibinfo
  {volume} {98}},\ \bibinfo {pages} {042702} (\bibinfo {year}
  {2018})}\BibitemShut {NoStop}%
\bibitem [{\citenamefont {Morita}\ \emph {et~al.}(2017)\citenamefont {Morita},
  \citenamefont {K\l{}os}, \citenamefont {Buchachenko},\ and\ \citenamefont
  {Tscherbul}}]{Morita_PRA17}%
  \BibitemOpen
  \bibfield  {author} {\bibinfo {author} {\bibfnamefont {M.}~\bibnamefont
  {Morita}}, \bibinfo {author} {\bibfnamefont {J.}~\bibnamefont {K\l{}os}},
  \bibinfo {author} {\bibfnamefont {A.~A.}\ \bibnamefont {Buchachenko}}, \ and\
  \bibinfo {author} {\bibfnamefont {T.~V.}\ \bibnamefont {Tscherbul}},\ }\href
  {\doibase 10.1103/PhysRevA.95.063421} {\bibfield  {journal} {\bibinfo
  {journal} {Phys. Rev. A}\ }\textbf {\bibinfo {volume} {95}},\ \bibinfo
  {pages} {063421} (\bibinfo {year} {2017})}\BibitemShut {NoStop}%
\bibitem [{SM()}]{SM}%
  \BibitemOpen
  \href@noop {} {}\bibinfo {note} {See Supplemental Material at
  [http://link.aps.org/ supplemental/] for a brief description of the quantum
  scattering methodology, values of physical constants and numerical parameters
  used in our CC calculations.}\BibitemShut {Stop}%
\bibitem [{\citenamefont {Tscherbul}\ and\ \citenamefont
  {Dalgarno}(2010)}]{Tscherbul_JPC10}%
  \BibitemOpen
  \bibfield  {author} {\bibinfo {author} {\bibfnamefont {T.~V.}\ \bibnamefont
  {Tscherbul}}\ and\ \bibinfo {author} {\bibfnamefont {A.}~\bibnamefont
  {Dalgarno}},\ }\href {\doibase 10.1063/1.3503500} {\bibfield  {journal}
  {\bibinfo  {journal} {The Journal of Chemical Physics}\ }\textbf {\bibinfo
  {volume} {133}},\ \bibinfo {pages} {184104} (\bibinfo {year}
  {2010})}\BibitemShut {NoStop}%
\bibitem [{\citenamefont {Gonz\'alez-Mart\'{\i}nez}\ and\ \citenamefont
  {Hutson}(2013)}]{Hutson_PRL13}%
  \BibitemOpen
  \bibfield  {author} {\bibinfo {author} {\bibfnamefont {M.~L.}\ \bibnamefont
  {Gonz\'alez-Mart\'{\i}nez}}\ and\ \bibinfo {author} {\bibfnamefont {J.~M.}\
  \bibnamefont {Hutson}},\ }\href {\doibase 10.1103/PhysRevLett.111.203004}
  {\bibfield  {journal} {\bibinfo  {journal} {Phys. Rev. Lett.}\ }\textbf
  {\bibinfo {volume} {111}},\ \bibinfo {pages} {203004} (\bibinfo {year}
  {2013})}\BibitemShut {NoStop}%
\bibitem [{\citenamefont {Kozyryev}\ \emph {et~al.}(2017)\citenamefont
  {Kozyryev}, \citenamefont {Baum}, \citenamefont {Matsuda}, \citenamefont
  {Augenbraun}, \citenamefont {Anderegg}, \citenamefont {Sedlack},\ and\
  \citenamefont {Doyle}}]{Kozyryev:17}%
  \BibitemOpen
  \bibfield  {author} {\bibinfo {author} {\bibfnamefont {I.}~\bibnamefont
  {Kozyryev}}, \bibinfo {author} {\bibfnamefont {L.}~\bibnamefont {Baum}},
  \bibinfo {author} {\bibfnamefont {K.}~\bibnamefont {Matsuda}}, \bibinfo
  {author} {\bibfnamefont {B.~L.}\ \bibnamefont {Augenbraun}}, \bibinfo
  {author} {\bibfnamefont {L.}~\bibnamefont {Anderegg}}, \bibinfo {author}
  {\bibfnamefont {A.~P.}\ \bibnamefont {Sedlack}}, \ and\ \bibinfo {author}
  {\bibfnamefont {J.~M.}\ \bibnamefont {Doyle}},\ }\href {\doibase
  10.1103/PhysRevLett.118.173201} {\bibfield  {journal} {\bibinfo  {journal}
  {Phys. Rev. Lett.}\ }\textbf {\bibinfo {volume} {118}},\ \bibinfo {pages}
  {173201} (\bibinfo {year} {2017})}\BibitemShut {NoStop}%
\bibitem [{\citenamefont {Liu}\ \emph {et~al.}(2017)\citenamefont {Liu},
  \citenamefont {Vashishta}, \citenamefont {Djuricanin}, \citenamefont {Zhou},
  \citenamefont {Zhong}, \citenamefont {Mittertreiner}, \citenamefont {Carty},\
  and\ \citenamefont {Momose}}]{Momose_PRL17}%
  \BibitemOpen
  \bibfield  {author} {\bibinfo {author} {\bibfnamefont {Y.}~\bibnamefont
  {Liu}}, \bibinfo {author} {\bibfnamefont {M.}~\bibnamefont {Vashishta}},
  \bibinfo {author} {\bibfnamefont {P.}~\bibnamefont {Djuricanin}}, \bibinfo
  {author} {\bibfnamefont {S.}~\bibnamefont {Zhou}}, \bibinfo {author}
  {\bibfnamefont {W.}~\bibnamefont {Zhong}}, \bibinfo {author} {\bibfnamefont
  {T.}~\bibnamefont {Mittertreiner}}, \bibinfo {author} {\bibfnamefont
  {D.}~\bibnamefont {Carty}}, \ and\ \bibinfo {author} {\bibfnamefont
  {T.}~\bibnamefont {Momose}},\ }\href {\doibase
  10.1103/PhysRevLett.118.093201} {\bibfield  {journal} {\bibinfo  {journal}
  {Phys. Rev. Lett.}\ }\textbf {\bibinfo {volume} {118}},\ \bibinfo {pages}
  {093201} (\bibinfo {year} {2017})}\BibitemShut {NoStop}%
\bibitem [{\citenamefont {Miller}(1970)}]{Miller:70}%
  \BibitemOpen
  \bibfield  {author} {\bibinfo {author} {\bibfnamefont {W.~H.}\ \bibnamefont
  {Miller}},\ }\href@noop {} {\bibfield  {journal} {\bibinfo  {journal} {J.
  Chem. Phys.}\ }\textbf {\bibinfo {volume} {52}},\ \bibinfo {pages} {543}
  (\bibinfo {year} {1970})}\BibitemShut {NoStop}%
\bibitem [{\citenamefont {Croft}\ \emph {et~al.}(2017)\citenamefont {Croft},
  \citenamefont {Makrides}, \citenamefont {Li}, \citenamefont {Petrov},
  \citenamefont {Kendrick}, \citenamefont {Balakrishnan},\ and\ \citenamefont
  {Kotochigova}}]{Croft:17}%
  \BibitemOpen
  \bibfield  {author} {\bibinfo {author} {\bibfnamefont {J.~F.~E.}\
  \bibnamefont {Croft}}, \bibinfo {author} {\bibfnamefont {C.}~\bibnamefont
  {Makrides}}, \bibinfo {author} {\bibfnamefont {M.}~\bibnamefont {Li}},
  \bibinfo {author} {\bibfnamefont {A.}~\bibnamefont {Petrov}}, \bibinfo
  {author} {\bibfnamefont {B.~K.}\ \bibnamefont {Kendrick}}, \bibinfo {author}
  {\bibfnamefont {N.}~\bibnamefont {Balakrishnan}}, \ and\ \bibinfo {author}
  {\bibfnamefont {S.}~\bibnamefont {Kotochigova}},\ }\href@noop {} {\bibfield
  {journal} {\bibinfo  {journal} {Nature Communications}\ }\textbf {\bibinfo
  {volume} {8}},\ \bibinfo {pages} {15897} (\bibinfo {year}
  {2017})}\BibitemShut {NoStop}%
\bibitem [{\citenamefont {Frisch}\ \emph {et~al.}(2014)\citenamefont {Frisch},
  \citenamefont {Mark}, \citenamefont {Aikawa}, \citenamefont {Ferlaino},
  \citenamefont {Bohn}, \citenamefont {Makrides}, \citenamefont {Petrov},\ and\
  \citenamefont {Kotochigova}}]{Frisch:14}%
  \BibitemOpen
  \bibfield  {author} {\bibinfo {author} {\bibfnamefont {A.}~\bibnamefont
  {Frisch}}, \bibinfo {author} {\bibfnamefont {M.}~\bibnamefont {Mark}},
  \bibinfo {author} {\bibfnamefont {K.}~\bibnamefont {Aikawa}}, \bibinfo
  {author} {\bibfnamefont {F.}~\bibnamefont {Ferlaino}}, \bibinfo {author}
  {\bibfnamefont {J.~L.}\ \bibnamefont {Bohn}}, \bibinfo {author}
  {\bibfnamefont {C.}~\bibnamefont {Makrides}}, \bibinfo {author}
  {\bibfnamefont {A.}~\bibnamefont {Petrov}}, \ and\ \bibinfo {author}
  {\bibfnamefont {S.}~\bibnamefont {Kotochigova}},\ }\href {\doibase
  10.1038/nature13137} {\bibfield  {journal} {\bibinfo  {journal} {Nature}\
  }\textbf {\bibinfo {volume} {507}},\ \bibinfo {pages} {475} (\bibinfo {year}
  {2014})}\BibitemShut {NoStop}%
\bibitem [{\citenamefont {Croft}\ and\ \citenamefont {Bohn}(2014)}]{Croft:14}%
  \BibitemOpen
  \bibfield  {author} {\bibinfo {author} {\bibfnamefont {J.~F.~E.}\
  \bibnamefont {Croft}}\ and\ \bibinfo {author} {\bibfnamefont {J.~L.}\
  \bibnamefont {Bohn}},\ }\href {\doibase 10.1103/PhysRevA.89.012714}
  {\bibfield  {journal} {\bibinfo  {journal} {Phys. Rev. A}\ }\textbf {\bibinfo
  {volume} {89}},\ \bibinfo {pages} {012714} (\bibinfo {year}
  {2014})}\BibitemShut {NoStop}%
\bibitem [{\citenamefont {Mayle}\ \emph {et~al.}(2013)\citenamefont {Mayle},
  \citenamefont {Qu\'em\'ener}, \citenamefont {Ruzic},\ and\ \citenamefont
  {Bohn}}]{Mayle:13}%
  \BibitemOpen
  \bibfield  {author} {\bibinfo {author} {\bibfnamefont {M.}~\bibnamefont
  {Mayle}}, \bibinfo {author} {\bibfnamefont {G.}~\bibnamefont {Qu\'em\'ener}},
  \bibinfo {author} {\bibfnamefont {B.~P.}\ \bibnamefont {Ruzic}}, \ and\
  \bibinfo {author} {\bibfnamefont {J.~L.}\ \bibnamefont {Bohn}},\ }\href
  {\doibase 10.1103/PhysRevA.87.012709} {\bibfield  {journal} {\bibinfo
  {journal} {Phys. Rev. A}\ }\textbf {\bibinfo {volume} {87}},\ \bibinfo
  {pages} {012709} (\bibinfo {year} {2013})}\BibitemShut {NoStop}%
\end{thebibliography}%
\end{document}